\documentstyle[12pt]{article}
\begin{document}

\input{epsf}

\def\beq{\begin{equation}}
\def\eq{\end{equation}}
\def\beqa{\begin{eqnarray}}
\def\eqa{\end{eqnarray}}
\def\bars{\begin{eqnarray*}}
\def\ears{\end{eqnarray*}}

\newcommand{\be}{\begin{equation}}
\newcommand{\ee}{\end{equation}}
\newcommand{\lesssim}{\raisebox{-0.5ex}{$\stackrel{<}{\sim}$}}
\newcommand{\la}{\mbox{$\lambda$}}
\newcommand{\gh}{\mbox{$\gamma$}}
\newcommand{\ep}{\mbox{$\epsilon$}}
\newcommand{\vep}{\mbox{$\varepsilon$}}
\newcommand{\lb}{\mbox{$\underline {l}$}}
\newcommand{\xb}{\mbox{$\underline {x}$}}
\newcommand{\rb}{\mbox{$\underline {r}$}}
\newcommand{\epsilonb}{\mbox{$\underline {\epsilon}$}}
\newcommand{\no}{\noindent}

\title{Proton structure functions in the dipole picture of BFKL dynamics.}
\author{{\bf H. Navelet$^1$}\\
{\bf R. Peschanski$^1$}\\
{\bf Ch. Royon$^2$}\\
{\bf S. Wallon$^1$}\\
 \\
{\small 1- Service de Physique Th\'eorique, Centre d'Etudes de Saclay}\\
{\small F-91191 Gif-sur-Yvette  Cedex, FRANCE}\\
{\small 2- DAPNIA - SPP, Centre d'Etudes de Saclay}\\
{\small F-91191 Gif-sur-Yvette  Cedex, FRANCE}}

\date{}
\maketitle

\begin{abstract}
The $F_2$, $F_G$, $R = F_L/F_T$ proton structure functions are derived in the QCD dipole picture. Assuming $k_T$ and renormalization-group factorization, we relate deep-inelastic proton scattering to deep-inelastic onium scattering.  
We get a three-parameter fit of the 1994 H1 data in the low-$x,$ moderate $Q^2$
range. The ratios $F_G/F_2$ and $R$ are predicted without further adjustment.
The dipole picture of BFKL dynamics is shown to provide a relevant model for quantitatively describing the proton structure functions at HERA. The
predictions for $F_2$ and $F_G$ are compatible with the next-to-leading order
DGLAP analysis, while $R$ is expected to be significantly lower at very small $x$.
\end{abstract}
\vspace{5cm}
\noindent
\hspace{1cm}May 1996\hfill\\
\hspace*{1cm}SPhT T96/043\hfill\\
\hspace*{1cm}DAPNIA/SPP 96-08\hfill\\
\hspace*{1cm}hep-ph/96
\thispagestyle{empty}
\newpage
\setcounter{page}{1}

{\bf 1. Introduction}
\\

The purpose of the present paper is to show that the QCD dipole picture \cite{mueller1,mueller2,mueller3,nik} provides
a pertinent model for describing the proton structure functions at HERA in the low-$x$ and moderate $Q^2$ range.
Indeed the HERA experimental data on the proton structure function and particularly the most recent 1994 published data \cite{h1} give a good opportunity to check the validity of different QCD-inspired theoretical models.
The QCD dipole model contains the physics of the BFKL Pomeron \cite{lip}, i.e. the  leading $\ln 1/x$ resummation of the perturbative expansion of QCD.
In order to apply this model to proton structure function phenomenology, we shall use the $k_T$-factorization properties \cite{catani} in the context of the dipole model. In this scheme we give predictions for the whole set of $F_2, F_G$
and $R = F_L/F_T$ proton structure functions. $F_2$ is well reproduced with only
three parameters and the ratios $F_G/F_2$ and $R$ are predicted without any further adjustment.

To apply consistently the QCD dipole model to deep-inelastic scattering on a proton, we are led to introduce massive $q\bar{q}$ (onium) configurations inside the proton. On the one hand, we compute the coupling of gluons to the dipole
states in an onium (see formulae (\ref{vg},\ref{mellinv})). On the other hand we exhibit factorization constraints 
on the non-perturbative probabilities of finding an onium into the proton (see formula (\ref{wanom})).

In section {\bf 2}, we derive the expression for the onium structure function, including the
determination of the gluon-dipole coupling using $k_T$-factorization.
In section {\bf 3}, we use once more the $k_T$-factorization properties to get the predictions for the set of proton structure functions  $F_T,  F_L$ and $F_G.$ They are obtained as inverse Mellin transforms which can be calculated by the steepest-descent method. In section {\bf 4}, we adjust the 3 free parameters of the resulting expressions with the 1994 data on $F_2,$ and give the predictions for
$F_G$ and $R$ without further adjustment. The summary of results is in section {\bf 5}.
\\

{\bf 2. Onium structure functions}
\\

To get the structure function $F_2$ we rely on the $k_T$-dependent
factorization properties \cite{catani} which are valid at high energy (small $x$).
In a first step, we calculate the deep-inelastic cross-section $\sigma^{onium}$ of a photon of 
virtuality $Q^2$ on an onium state. This onium state can be described by  dipole configurations \cite{mueller1}.
The photon-onium cross-section reads

\beq
\label{oniumphoton}
\sigma^{onium} = \int^{ }_{ } d^2 \rb d z \Phi^{(0)} (\rb,z) \sigma(x,Q^2;r)
\eq

\no
where $\Phi^{(0)} (\rb,z)$ is the probability distribution of dipole 
configurations of transverse coordinate $\rb$ (with modulus $r$).

In the $k_T$-factorization scheme, one writes
\beq
\label{fact}
Q^2 \sigma (x,Q^2;r) = \int^{Q^2}_{ } d^2 \vec{k} \int^{1}_{0} 
\frac{d z}{z} \, \hat{\sigma} (x/z,\vec{k}^2 / Q^2) \, F(z,(kr)^2)
\eq

\noindent
where $\hat{\sigma}/Q^2$ is the $(\gamma~ g(k) \rightarrow q ~ \bar{q})$ Born cross section for an off-shell gluon of transverse momentum
$\vec{k}$. $F(z,(kr)^2)$ is the unintegrated gluon distribution of an onium state of size $r$  and contains the physics of the BFKL pomeron.
In the $x$-space, eq.(\ref{fact}) is a convolution . Using the inverse-Mellin transform 
\beq
\label{mellinx}
f(x) = \frac{1}{2i\pi} \int^{c+i\infty}_{c-i\infty}d\omega ~x^{-\omega}
f(\omega)
\eq
yields
\beq
\label{factmellinx}
Q^2 \sigma (x,Q^2;r) = \frac{1}{2i\pi} \int^{ }_{ } d \omega \int^{ }_{ } d^2 \vec{k}~ 
 \hat{\sigma} (\omega,\frac{\vec{k}^2}{Q^2}) F (\omega,(kr)^2) x^{-\omega}.
\eq

\no
A second Mellin transform in the ${k}^2$-space, namely
\beq
\label{mellinq}
f(\omega,(kr)^2) = \frac{1}{2i\pi} \int^{c+i\infty}_{c-i\infty} d \gamma
f(\omega,\gamma) (kr)^{2 \gamma} 
\eq

\no
yields
\beq
\label{factmellinxq}
Q^2 \sigma^ {\gamma} (x,Q^2;r) =  \int^{ }_{ } \frac{d \omega}{2i\pi} 
\int^{ }_{ }  \frac{d \gamma}{2i\pi}  
\hat{\sigma} (\omega,1 \!- \! \gamma) F(\omega,\gamma) (Qr)^{2 \gamma} x^{-\omega}.
\eq
 
\no
Here we have used the Mellin transform of both $\hat{\sigma}$ and $F$,
and taken into account the relation 
\beq
\frac{1}{2i\pi} \int^{ }_{ } \frac{d \vec{k}^2}{\vec{k}^2} 
(\vec{k}^2)^{-\gamma - \gamma' + 1} = \delta(\gamma' - (1-\gamma)).
\eq

The unintegrated gluon distribution for an onium of size $r$ has been derived 
in the QCD dipole picture. One obtains \cite{mueller1} 
\beq
\label{seckt}
F(\omega,\gamma) = \frac{\bar{\alpha} N_c}{\pi} \frac{1}{\gamma} \,
\frac{v(\gamma)}{\omega - \frac{\bar{\alpha} N_c}{\pi} \chi(\gamma)}
\eq
where $\chi(\gamma) = 2 \Psi(1) - \Psi(\gamma) - \Psi(1 - \gamma)$ with $\Psi(\gamma) = \frac{d \ln \Gamma}{d \gamma},$ and $\frac{\bar{\alpha }N_c}{\pi} \chi(\gamma)$ is the well-known BFKL
\cite{lip} anomalous dimension.  $v(\gamma)$ is, up to a factor, the Mellin transform of the gluon-dipole
coupling, which expression in the QCD dipole framework we shall now give. 

Using once more the $k_T$-factorization in extracting a gluon
of virtuality $l$ from a dipole of transverse radius $r$, we find the
following expression of $v$ in the momentum space \cite {mueller1,mueller2}
\beq
\label{vg}
v(rl) =  \frac{1}{4 \pi} \int^{2 \pi}_{0} d \theta \,(2 - e^{i l r \cos \theta} - e^{- i l r \cos \theta}) \equiv 1 - J_0(lr)
\eq
and correspondingly
\beq
\label{mellinv}
v(\gamma) = \int^{ }_{ } v(lr) \, (lr)^{-2 \gamma - 1} d(lr)
= \frac{2^{-2 \gamma - 1}}{\gamma} \frac{\Gamma(1-\gamma)}{\Gamma(1+\gamma)}.
\eq
A sketchy proof of the determination of $v(lr)$ is obtained by using the eikonal coupling of the
off-shell gluon to the dipole of transverse size $r$. We use the well-known light-cone coordinates as well as the light-cone gauge. 
When the emitted gluon is soft, the dominant component of its polarization \cite{mueller1} is 
\beq
\label{pol}
\epsilon^{\lambda}_- = \frac{\lb_.\epsilonb^{\lambda}}{l_{+}}
\eq
where $\epsilonb^{\lambda}$ is a two dimensional vector depending on the polarization index $\lambda$.
Given a quark (antiquark) of transverse
coordinate $\xb_0$ ($\xb_1$), the coupling has the form 
$J_{\mu} \epsilon^{\mu \lambda} \approx J_+ \epsilon_{- \lambda}$ with the
corresponding
eikonal current given by
\beq
\label{current}
J_+ = i g T^a (e^{-i \lb. \xb_1} - e^{-i \lb. \xb_0}) \frac{2 l_+}{\lb^2},
\eq
where $\bar{\alpha} = g^2/ (4 \pi)$.
\no
After averaging over color and polarization, this leads to the cross section integrand
\beqa
\label{cross}
\nonumber 
\frac{1}{N_c} Tr|J_+ \epsilon^{\lambda}_-|^2 &= \frac{1}{N_c} \sum_{\lambda=1,2} g^2 Tr(T^a T^a) (e^{-i \lb. \xb_1} - e^{-i \lb. \xb_0})
(e^{i \lb. \xb_1} - e^{i \lb. \xb_0}) 4 (\frac{\lb.\epsilonb^{\lambda}}{\lb^2})^2 \\
&= g^2 C_F(2 - e^{i \lb. \xb_{10}} - e^{- i \lb. \xb_{10}}) \frac{4}{\lb^2}.
\eqa
After integrating over the polar angle $\theta,$ and with the suitable normalization \cite{catani}, one gets formula (\ref{vg}).
Note that a similar expression is encountered in the elementary dipole-dipole cross section \cite{mueller2}.

Since $F(\omega,\gamma)$ (formula (\ref{seckt})) exhibits a pole at 
$\omega_p = \frac{\alpha_S N_c}{\pi} \chi(\gamma)$, the $\omega$ integral yields
\beq
\label{omegaint}
Q^2 \sigma (x,Q^2;r) = \frac{\bar{\alpha} N_c}{\pi} \int^{ }_{ } \frac{d \gamma}{2 i \pi}  
 \hat{\sigma} (\omega = \omega_p,1\!-\!\gamma) \frac{v(\gamma)}{\gamma} (r^2Q^2)^{\gamma} 
e^{\frac{\bar{\alpha} N_c}{\pi} \chi(\gamma) \ln(\frac{1}{x})}.
\eq
Furthermore 
$\hat{\sigma} (\omega,1\!-\!\gamma)$ is weakly $\omega$-dependent \cite{catani}, and thus 
\beq
\label{sigmah}
\hat{\sigma} (\omega_p,1\!-\!\gamma)~ \simeq  ~ \hat{\sigma} (0,1\!-\!\gamma) \equiv 4 \pi^2 \alpha_{e.m.} h(\gamma),
\eq
factorizing out the electromagnetic coupling.

In order to average over the wave function of the onium state (see formula (\ref{oniumphoton})), one defines 
\beq
\label{raverage}
<r^2>^{\gamma} = \int^{ }_{ } \, d^2\rb \ (r^2)^{\gamma} dz ~ \Phi^{(0)}(\rb,z) = (M^2)^{-\gamma}
\eq
where $M^2$ is a {\it perturbative scale} characterizing the average onium size. This means that $M^2$ is  high enough  to use the perturbative dipole-cascade mechanism \cite{mueller1}.
The photon-onium structure function then reads

\beqa
\label{crossonium}
F^{onium} (x,Q^2/M^2) &=& \frac{2\bar{\alpha} N_c}{\pi} \int^{ }_{ } \frac{d \gamma}{2 i \pi} \left(\frac{Q^2}{M^2} \right )^{\gamma} 
 h(\gamma) \frac{v(\gamma)}{\gamma}  
e^{\frac{\bar{\alpha} N_c}{\pi} \chi(\gamma) \ln(\frac{1}{x})} \nonumber \\
&\equiv&  \int^{ }_{ } \frac{d \gamma}{2 i \pi} \left(\frac{Q^2}{M^2} \right )^{\gamma}F(x,\gamma),
\eqa
where $F(x,\gamma)$ is the inverse Mellin transform of the onium structure function.
\\

{\bf 3. Proton structure functions}
\\

Let us now deal with deep inelastic scattering on a proton target.
Assuming   the $k_T$-factorization properties to be valid for high-energy scattering off a  proton target,
we substitute in formula (\ref{crossonium}) $F(x,\gamma)$ by
$F(x,\gamma) w(\gamma,M^2;{Q_0}^2)$ where $w$ can be interpreted as the Mellin-transformed probability of finding an onium of transverse mass $M^2$ in the proton. ${Q_0}^2$ is a typically non-perturbative proton scale. 
Thus,
\beq
\label{crosshadron}
F^{proton}(x,Q^2;Q_0^2) = \frac{2\bar{\alpha} N_c}{\pi} \int^{ }_{ } \frac{d \gamma}{2 i \pi}  
 h(\gamma) \frac{v(\gamma)}{\gamma} \left(\frac{Q^2}{M^2}\right) ^{\gamma} 
e^{\frac{\bar{\alpha} N_c}{\pi} \chi(\gamma) \ln(\frac{1}{x})} w(\gamma,M^2;{Q_0}^2).
\eq

As previously mentionned, $M^2$ is an arbitrary perturbative scale. The overall result has to be independent of the factorization scale, provided it is in the perturbative region. Hence,   assuming the renormalization group properties to be valid, the $M^2$ dependence of $w$ has to match the $(M^2)^{-\gamma}$ dependence of the term $(\frac{Q^2}{M^2})^{\gamma}.$ 
One  then writes
\beq
\label{wanom}
w(\gamma,M^2;{Q_0}^2) 
=  w(\gamma)\ \left(\frac{{M}^2}{Q_0^2}\right)^{\gamma}. 
\eq
The argument leading to formula (\ref{wanom}) is similar to the one applied to the one-loop QCD evolution equation in ref. \cite{wil}  .
This yields the final result
\beqa
\label{crosshadronnonpert}
F^{proton}(x,Q^2;Q_0^2) &= 2 \frac{\bar{\alpha} N_c}{\pi} \int^{ }_{ } \frac{d \gamma}{2 i \pi}  
 h(\gamma) \frac{v(\gamma)}{\gamma}  w(\gamma) \left(\frac{Q^2}{Q_0^2}\right)^{\gamma} 
e^{\frac{\bar{\alpha} N_c}{\pi} \chi(\gamma) \ln(\frac{1}{x})}.
\eqa
Note that $F^{proton}$  depends on the ratio $Q^2/{Q_0}^2$ where ${Q_0}^2$ can be interpreted as a characteristic scale of   the non perturbative hadron-onium structure function.

Let us use the generic result (\ref{crosshadronnonpert}) to derive the prediction
for specific proton structure functions in the QCD dipole picture:
\beq
\label{predgen}
\left(\begin{array}{c}
F_T \\ F_L \\ F_G \end{array} \right) 
= \frac{2\bar{\alpha} N_c}{\pi} \int^{ }_{ }  \frac{d \gamma}{2 i \pi}  
  \left(\frac{Q^2}{{Q_0}^2}\right) ^{\gamma} 
e^{\frac{\bar{\alpha} N_c}{\pi} \chi(\gamma) \ln(\frac{1}{x})} \left(\begin{array}{c}
h_T \\ h_L \\ 1 \end{array} \right)
\frac{v(\gamma)}{\gamma} w(\gamma)
\eq
where $F_{T(L)}$ is the structure function corresponding to transverse (longitudinal) photons and $F_G$ the gluon structure function.
The coefficient functions 
\beq
\label{defh}
\left(\begin{array}{c}
h_T \\ h_L \end{array} \right) = \frac{\bar{\alpha}}{ 3 \pi \gamma} 
\frac{(\Gamma(1 - \gamma) \Gamma(1 + \gamma))^3}{\Gamma(2 - 2\gamma) \Gamma(2 + 2\gamma)} \frac{1}{1 - \frac{2}{3} \gamma} \left( \begin{array}{c} (1 + \gamma)
(1 - \frac{\gamma}{2}) \\ \gamma(1 - \gamma) \end{array} \right)
\eq
are given in ref \cite{catani}. Note that the coupling constant $\bar{\alpha}$
 in formula (\ref{defh}) is kept identical to the effective coupling of the BFKL kernel in formula (\ref{seckt}). In the following we shall take $\bar{\alpha}$ constant and not running, as in general considered in both BFKL and
$k_T$-factorization schemes. 
 
The integral in $\gamma$ is performed by the steepest descent method.  Assuming that $w(\gamma)$ is smooth and regular near $\gamma = \frac{1}{2},$ we can calculate the structure functions by expansion of the BFKL kernel near this value. The saddle point is at 
\beqa
\label{steep}
\nonumber
\gamma_c &=& \frac{1}{2} ( 1 - a \ln \frac{Q}{Q_0}) \\
a  &=& \left(\frac{\bar{\alpha} N_c}{\pi} 7 \zeta(3) \ln\frac{1}{x}\right) ^{-1}.
\eqa
Formulae (\ref{steep}) show that the considered approximation is valid when $$a \ln(\frac{Q}{Q_0}) \simeq \ln \frac{Q}{Q_0}/\ln\frac{1}{x} \ll 1,$$ that is the small $x$, moderate $Q/Q_0$ kinematical domain.

We obtain
\beq
\label{predF2}
F_2 \equiv F_T + F_L = C a^{1/2} e^{(\alpha_{P} -1) \ln\frac{1}{x}} \frac{Q}{Q_0} e^{- \frac{a}{2} \ln^2 \frac{Q}{Q_0}},
\eq
where
\beq
\label{alphap}
\alpha_{P} -1 = \frac{4 \bar{\alpha} N_{C} \ln 2}{\pi}
\eq
 $C,Q_0$ and $\alpha_{P}$ are three unknown parameters which have to be determined by the fit.
Once the $F_2$ fit performed, formula (\ref{crosshadronnonpert}) allows one to predict $F_G$ and $R = F_L / F_T$.
Moreover $R$  and $F_G / F_2$ are 
independent of the overall normalization $C$, i.e of the non-perturbative $w(\gamma_c)$.
One gets 
\beq
\label{fgf2}
\frac{F_G}{F_2} = \left. \frac1{h_T + h_L}\right|_{\gamma = \gamma_c}\equiv \frac{3 \pi \gamma_c}{\bar{\alpha}} \frac{1 - \frac{2}{3}\gamma_c}{1 + \frac{3 }{2}\gamma_c - \frac{3}{2}\gamma_c^2} \frac{\Gamma(2 - 2 \gamma_c) \Gamma(2 + 2 \gamma_c)}{(\Gamma(1 - \gamma_c)\Gamma(1 + \gamma_c))^3}  
\eq
and
\beq
\label{R}
R = \frac{h_L}{h_T}(\gamma_c) = \frac{\gamma_c (1 - \gamma_c)}{(1 + \gamma_c)(1 - \frac{\gamma_c}{2})}.
\eq

{\bf 4. Comparison with experiment and predictions}
\\

In order to test the accuracy of the $F_{2}$ 
parametrisation obtained in formula (\ref{predF2}),
a fit using
the recently published data from the H1 experiment \cite{h1} has been performed.
In order to remain in the domain of validity of the QCD dipole model,
all the points with $Q^{2} \leq 150$ $GeV^{2}$ were used in the fit (we
checked that the result is not strongly dependent on the cut value). The 
parameters used for the fit are $C$, $Q_{0}$, and $\alpha_{P}$. The fit for $F_{2}$
is displayed in figure 1. The $\chi^{2}$ is 101 for 130
points for   parameter values:
\begin{eqnarray}
&~& \alpha_{P}=1.282  \nonumber \\
&~& Q_{0}=0.627 \, GeV. \nonumber \\
&~& C=0.077
\end{eqnarray}
The value of the hard Pomeron intercept $\alpha_P$ is in agreement with values found
in other determinations using BFKL dynamics for describing $F_{2}$ at HERA \cite{agkms}.
It corresponds to an effective coupling constant $\bar{\alpha} \simeq
.11$. As a matter of comparison, the input value
corresponding to the H1 QCD study is $.12$. 
As discussed in section 3 of this paper, $Q_{0}$ and $C$ are  
parameters of non-perturbative origin. The value of $Q_0$, which corresponds to
an average transverse size of $.3$ fermi, is 
in the range expected for deep inelastic scattering on a proton. $C$ is connected to the non-perturbative input $w(1/2)$.

\par 
With only three
parameters, the fit of the data is remarkably good in the low $Q^{2}$ range. When considering the whole set of H1 data (see fig. 2), deviations from the fit are only observed in the high-$Q^2$ range ($Q^{2} \geq 200 \, GeV^{2}$), for $x > 10^{-2}$. This is expected since the dipole model does not contain the valence contribution and does not describe the high $Q^2$ QCD dynamics.  
By comparison with our previous
results  \cite{NPR} based on the H1 and Zeus 1993 data \cite{H1ZEUS},
the quality of the present fit confirms the validity of our predictions, while taking into account the extended kinematical domain and the more precise data taken in 1994.

\par
Using formula (\ref{fgf2}), we get now a precise prediction 
of the gluon density. The value of $\bar{\alpha}$ we used in formula (\ref{fgf2}) is the same as for $F_2$, that is the one extracted from the fitted value of $\alpha_P$ by the help of eq.(\ref{alphap}).
The comparison with the
published H1 determination \cite{h1} is shown in figure 3. The grey band is the gluon
value obtained by the H1 collaboration from their DGLAP next-to-leading order
QCD fit for two different values of $Q^{2}$. 
We can see that our prediction is 
quite satisfactory without any further adjustment for
the H1 values of $F_{G}$ for $x \leq 5.10^{-2}$
(this parametrisation is not supposed to be valid for  higher values 
of $x$). We notice that the fit is located in the  range quoted by
the H1 collaboration, while all the gluon density values situated in the error 
band are equiprobable \cite{h1,joel}. In figure 3, we have also shown
the result using only the first-order perturbative expansion of $h_T$ and $h_L$
in Eq.(\ref{defh}). 
This corresponds to taking
\begin{eqnarray}
\label{dll}
\frac{F_{G}}{F_{2}} = \left. \frac{3 \pi \gamma}{\bar{\alpha} (1 + \gamma)} \right|_{\gamma =\gamma_c}. 
\end{eqnarray} 
The comparison between Eqs. (\ref{fgf2}) and (\ref{dll}) exhibits the resummation effect of the $\ln 1/x$ terms on the coefficient functions
$h_{T}$ and $h_{L}$. The result of formula (\ref{dll}) is shown as a dotted line in fig.3. The H1 determination clearly favours the resummed result. Our results show
in particular that the gluon distribution does not allow
to make a distinction between the DGLAP equations, used for the H1 determination, and the BFKL dynamics expressed by the
QCD dipole model.   

\par
Let us comment the prediction of this model for the value of $R=F_{L}/F_{T}$
(formula \ref{R}). We first
notice that this prediction is independent of the non-perturbative
normalisation $C$ obtained in the fit of the structure function $F_{2}$. The only
non-perturbative parameter which enters the $R$ prediction is $Q_{0}$. The corresponding curve (full line) is displayed in 
figure 4. As before, it is useful to compare this prediction with the 
one-loop one (dotted line), namely $R= \gamma_{c}$.

The measurement of $R$ might be an
opportunity to distinguish between the BFKL and DGLAP mechanisms. One expects the Altarelli-Martinelli \cite{am} predictions for $R$ to be larger \cite{ad} than the limiting value $2/9$ obtained in the dipole model. Moreover, the estimated values quoted in ref.\cite{h1}, using the H1 determination of the gluon structure
function, are significantly higher than our predictions.
It is known that this measurement is very difficult to achieve and the $R$ values
are very much correlated to the gluon densities when a QCD fit is performed.
However, the fact that a sizeable difference for $R$ is predicted in the same region where $F_2$ and $F_G$ are well described is an incentive for further studies
of $R$.
\\

{\bf 5. Summary of results}
\\

The main results of our analysis of deep-inelastic scattering can be summarized 
as follows
\\

i) Using (twice) the $k_T$-factorization properties of high-energy hard interactions\cite{catani} we were able to give a generic formula for the different structure functions of the proton, and by straightforward extension to any physical target, namely
\beqa
\label{Ftarget}
F^{Target}(x,Q^2;Q_0^2) &= \frac{2\bar{\alpha} N_c}{\pi} \int^{ }_{ } \frac{d \gamma}{2 i \pi} \left(\frac{Q^2}{{Q_0}^2}\right)^{\gamma} 
 h(\gamma) \frac{v(\gamma)}{\gamma}   
e^{\frac{\bar{\alpha} N_c}{\pi} \chi(\gamma) \ln(\frac{1}{x})} w^{Target}(\gamma),
\eqa
where $h(\gamma)$ is associated \cite{catani} to the  $(\gamma~ g \rightarrow q ~ \bar{q})$ Born cross section (see (\ref {sigmah})), $v(\gamma)$ is the gluon-dipole coupling function (\ref{mellinv}) and $w^{Target}$ is the probability of finding an onium in the target.
\\

ii) Using the steepest-descent method, we obtain a 3-parameter expression for 
$F_2$ which gives a fair description of all the 1994 H1 data with $Q^2 \le 150 \ GeV^2.$ 
\\

iii) Without further adjustment, the model gives definite predictions for $F_G$ and $R.$ One obtains:
\be
\label{fgf2bis}
\frac{F_G}{F_2} = \left. \frac1{h_T + h_L}\right|_{\gamma = \gamma_c};\ 
 R = \left.\frac{h_L}{h_T}\right|_{\gamma = \gamma_c},
\eq
 where the expression of the saddle-point $\gamma_c$ is given in formula (\ref{steep}).
In particular, the resulting $F_G$ is fully compatible with the determination by H1\cite{h1}.
\\

iv) The predictions of the QCD dipole model are similar to those obtained through next-to-leading order DGLAP evolution equations but for $R$, which differs from the 
 predictions at very small $x$ using the Altarelli-Martinelli equation.

\bigskip
{\bf Acknowledgments}
We thank A. Bialas and J. Feltesse for fruitful discussions, and M. Chemtob for
a careful reading of the manuscript.
\eject

\eject

{\bf FIGURE CAPTIONS}
\\

{\bf Figure 1}
{\it 3-parameter fit of the H1 proton structure function for $Q^{2} \leq 150 \, GeV^{2}$}. The fit is constrained by
the 1994 data from the H1 experiment \cite{h1} (including the point at 2 $GeV^2$
not shown on the figure for convenience). 
\\

{\bf Figure 2}
{\it Comparison of the fit with all 1994 H1 data}. The validity of the fit extends beyond the domain included in the chi-squared (the highest $Q^2$ point at $5000 \ GeV^2$ is not displayed for convenience). We note a discrepancy
at high $x$, and high $Q^{2},$ as expected outside the validity of the model.
\\

{\bf Figure 3}
{\it Prediction for the gluon density $F_{G}(x,Q^{2})$.} We compare our prediction
(full line)  (cf. text) 
with the gluon density obtained by the H1 collaboration \cite{h1} (grey band) from their
DGLAP next-to-leading order QCD fit. In order to exhibit the resummation effect, we also display the one loop approximation (dotted line), cf. formula (\ref{dll}).
\\

{\bf Figure 4}
{\it Prediction for $R$.} The full line describes our
prediction, and is compared to the one-loop approximation
(dotted line). 

\eject
\input epsf
\vsize=25.truecm
\hsize=18.truecm

\epsfxsize=18.cm{\centerline{\epsfbox{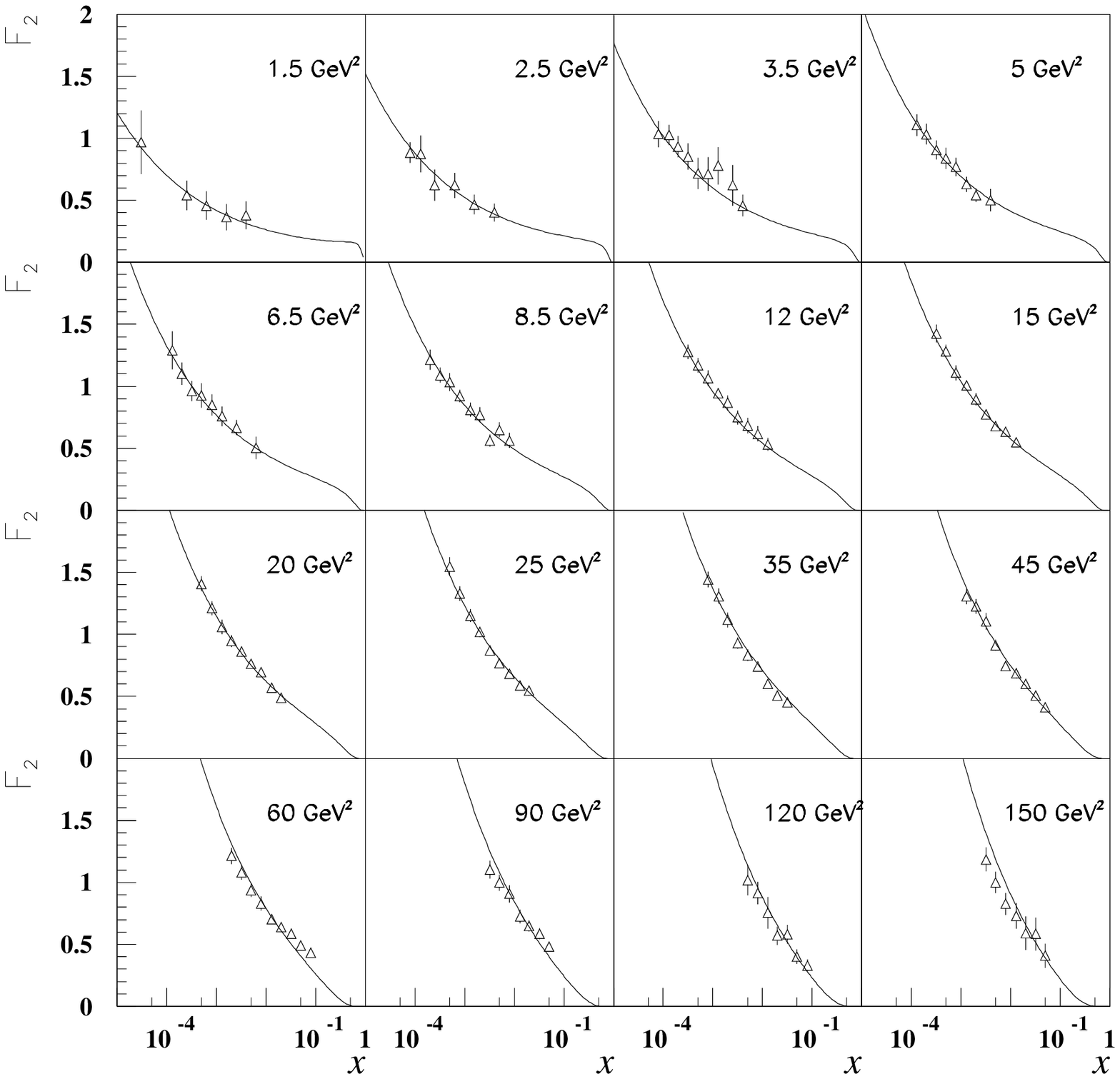}}}
\eject

\vsize=25.truecm
\hsize=18.truecm
\epsfxsize=18.cm{\centerline{\epsfbox{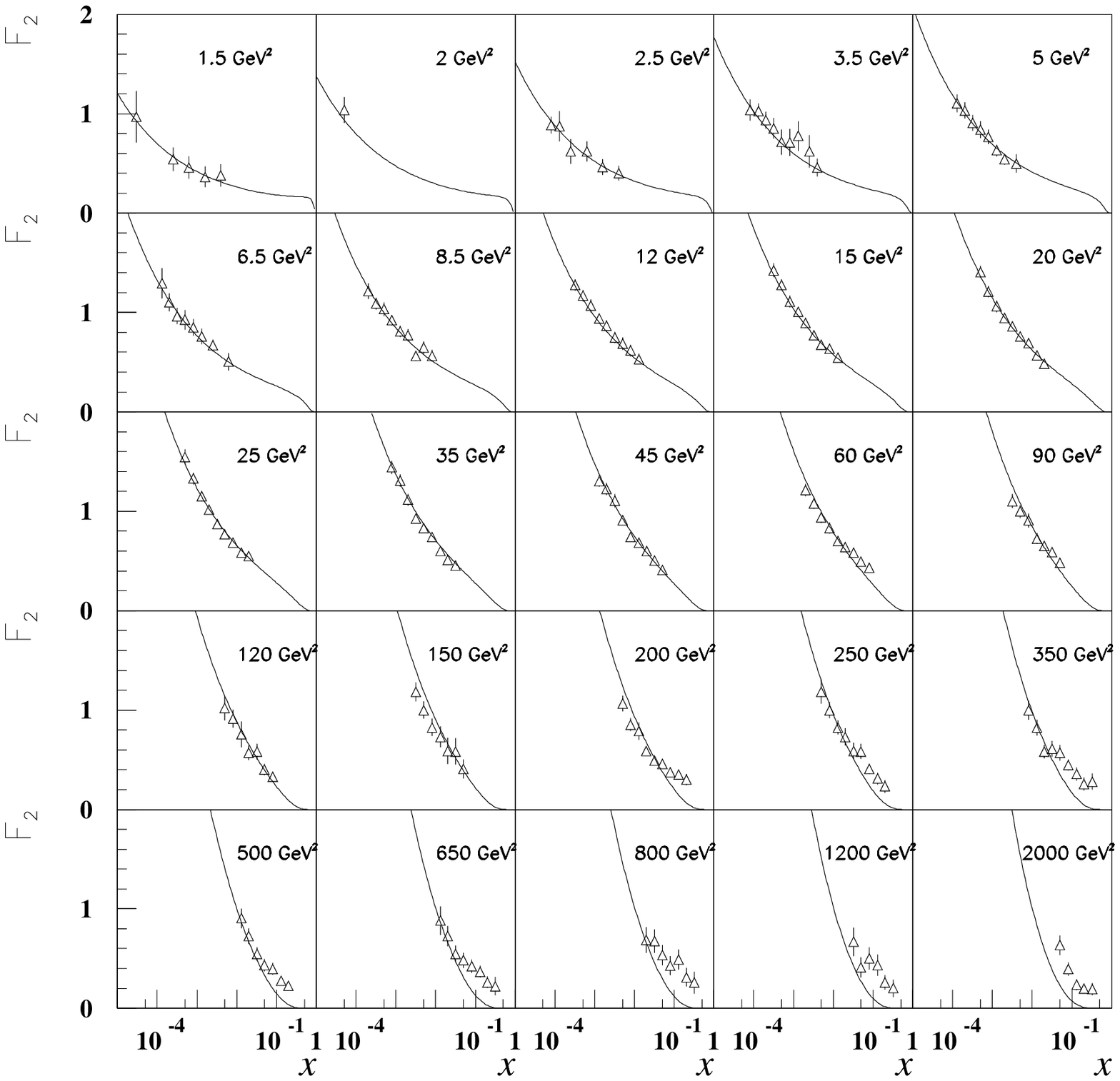}}}
\eject

\vsize=25.truecm
\hsize=18.truecm
\epsfxsize=18.cm{\centerline{\epsfbox{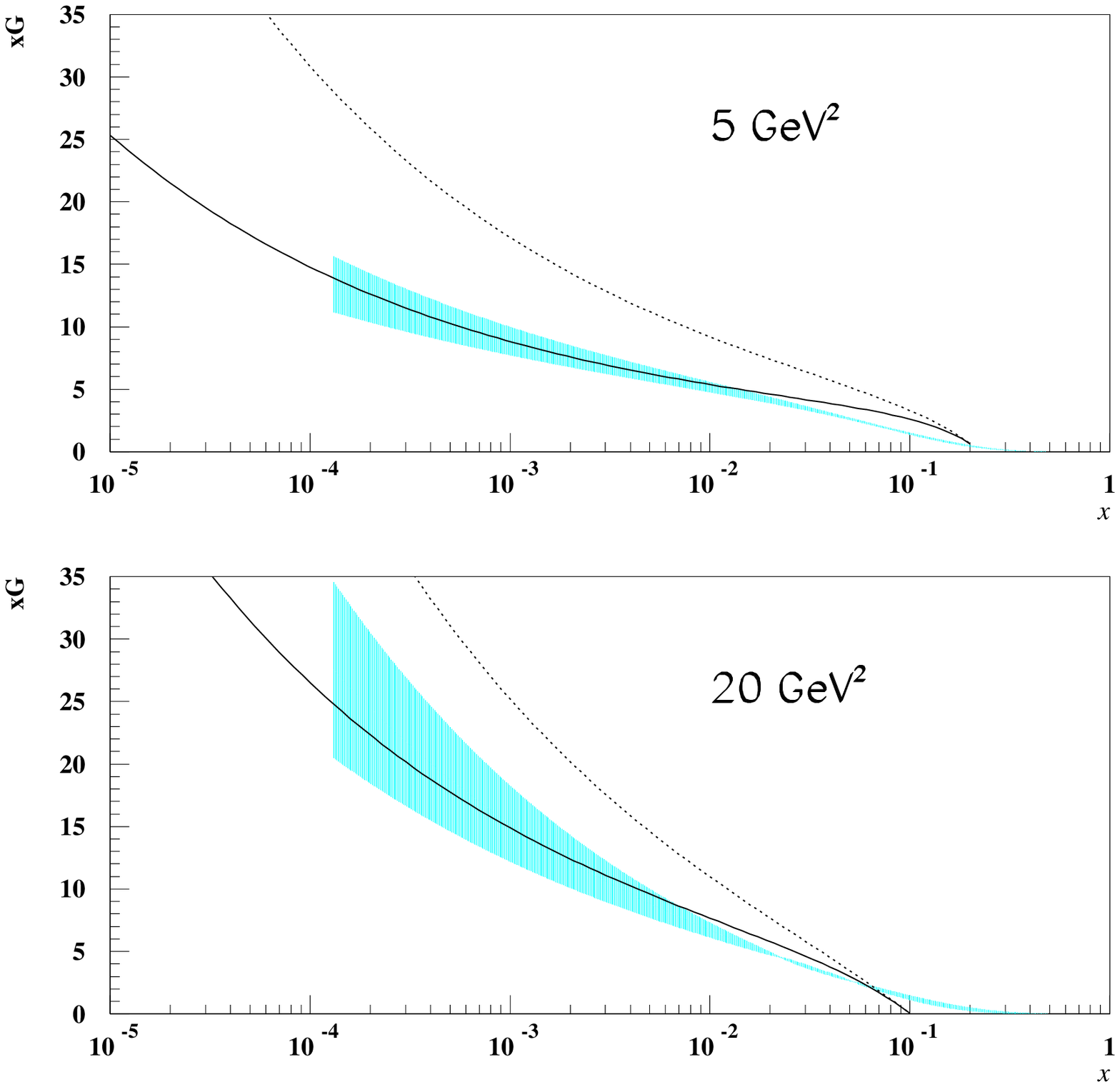}}}
\eject

\vsize=25.truecm
\hsize=18.truecm
\epsfxsize=18.cm{\centerline{\epsfbox{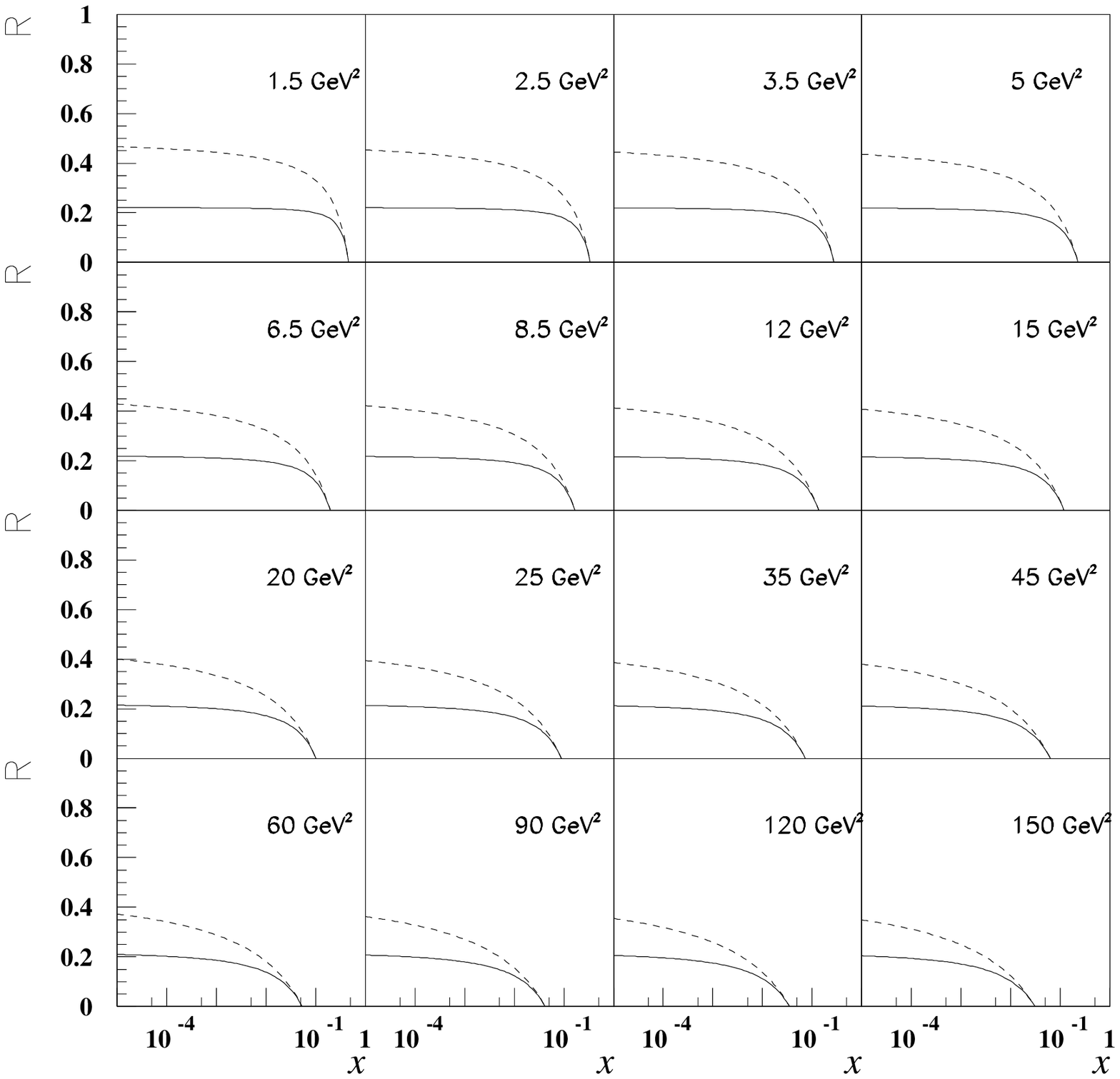}}}
\eject


\begin{thebibliography}{90}

\bibitem{mueller1}
A.H.Mueller, {\it Nucl. Phys.} {\bf B415} (1994) 373.

\bibitem{mueller2}
A.H.Mueller and B.Patel, {\it Nucl. Phys.} {\bf B425} (1994) 471.

\bibitem{mueller3}
A.H.Mueller, {\it Nucl. Phys.} {\bf B437} (1995) 107.

\bibitem{nik}
N.N.Nikolaev and B.G.Zakharov, {\it Zeit. f\"ur. Phys.} {\bf C49} (1991) 607;
{\it ibid.} {\bf C64} (1994) 651.

\bibitem{h1}
H1 coll., S.Aid et al. preprint DESY 96-039, March 1996.

\bibitem{lip}
V.S.Fadin, E.A.Kuraev and L.N.Lipatov   {\it Phys. Lett.} {\bf B60} (1975)
50; I.I.Balitsky and L.N.Lipatov, {\it Sov.J.Nucl.Phys.} {\bf 28} (1978) 822.

\bibitem{catani}
S.Catani, M.Ciafaloni and Hautmann, {\it Phys. Lett.} {\bf B242} (1990) 97;
{\it Nucl. Phys.} {\bf B366} (1991) 135; J.C.Collins and R.K.Ellis, {\it Nucl. Phys.} {\bf B360} (1991) 3; S.Catani and Hautmann, {\it Phys. Lett.} {\bf B315} (1993) 157;
{\it Nucl. Phys.} {\bf B427} (1994) 475

\bibitem{H1ZEUS}
H1 coll., T.Ahmed et al. \ {\it Phys.Lett.} {\bf B348} (1995) 681;
ZEUS coll., M.Derrick et al. {\it Zeit. f\"ur. Phys.} {\bf C68} (1995) 569. 

\bibitem{joel}
J. Feltesse, private communication.



\bibitem{NPR}
H.Navelet, R.Peschanski and Ch.Royon, {\it Phys. Lett.} {\bf B366} (1995) 329.


\bibitem{dglap}G. Altarelli and G. Parisi, {\it Nucl. Phys.} {\bf B126} (1977)
298; V.N. 
Gribov and L.N. Lipatov, {\it Sov. Journ. Nucl. Phys.} {\bf 15} (1972) 438 and
675.

\bibitem{wil}A. De R\'ujula, S. L. Glashow, H. D. Politzer, S. B. Treiman, F. Wilczek, and A. Zee, {\it Phys. Rev.} {\bf D10} {1974} 1649.


\bibitem{agkms}A.J. Askew, K. Golec-Biernat, J. Kwieci\' nski, A.D. Martin
and P.J. 
Sutton, {\it Phys. Lett.} {\bf B325} (1994) 212.

\bibitem{am}G. Altarelli and G. Martinelli, {\it Phys. Lett.} {\bf B76} (1978) 89.

\bibitem{ad}A. Dyring (NMC coll.), PhD Thesis, Uppsala 1995.



\end{thebibliography}
\end{document}